\documentclass[11pt,a4paper]{article}
\usepackage{jheppub}
\usepackage{amsmath,xfrac,xcolor,amssymb,float,cancel}

\title{The $D3$-${\bar{D3}}$-Brane Inflation Model Revisited}

\author[]{S.-H. Henry Tye}

\affiliation[]{Department of Physics,
Cornell University, Ithaca, NY 14853, USA}

\affiliation[]{Department of Physics,
Hong Kong University of Science and Technology, Clear Water Bay, \\
Hong Kong, SAR, China}

\emailAdd{tye.henry@gmail.com}

	\abstract{Combining previous results, the $D3$-${\bar{D3}}$-brane pair potential $U(T, \phi)$ is presented here, where the inflaton $\phi$ measures the separation between the $D3$-brane and the anti-${D3}$-brane, and the complex scalar mode $T$ becomes tachyonic when the annihilation of the branes happens as they collide. Besides the distinct form of the inflationary potential, this hybrid inflationary model differs from a typical hybrid model in 2 important aspects: (1) $U(T, \phi)$ becomes complex when $T$ becomes tachyonic, where ${\rm{Im}}\,U(T, \phi)$ plays an important role in the dynamics towards the end of the inflationary epoch; (2) tunnelling during the inflationary epoch can happen; this is particularly relevant if there are multiple $D3$-${\bar{D3}}$-brane pairs in different warped throats. Besides the production of cosmic superstrings, the model offers the possibility of first order phase transition that may generate large enough density perturbation for the primordial black hole production. Stochastic gravitational wave background from these sources remain to be fully investigated. 
		}

\begin{document}
	
	\maketitle

\section{Introduction}
	
	It is generally believed that our universe started with a cosmic inflationary epoch~\cite{Guth:1980zm,Linde:1981mu}. In the brane world scenario in string theory, where we live in a stack of $3$-branes (spanning the 3 large dimensions) inside the 9 dimensional space (where the remaining 6 dimensions are dynamically compactified), brane inflation is quite natural~\cite{Dvali:1998pa}.  The simplest and most natural scenario is the $D3$-brane and the anti-$D3$(${\bar{D3}}$)-brane inflation in Type IIB string theory~\cite{Dvali:2001fw,Burgess:2001fx}.
However, its toy model version does not work, as the attractive force between the $D3$-brane and the ${\bar{D3}}$-brane is too strong in the limited compactified space~\cite{Burgess:2001fx}. On the other hand, putting the $D3$-${\bar{D3}}$-brane pair in a more realistic setting works beautifully : dynamical compactification of the 6 extra spatial dimensions requires warped geometry~\cite{Giddings:2001yu}, and thus the presence of warped throats. The $D3$-branes and the ${\bar D3}$-branes naturally settle in
warped throats and it is the warping that weakens the attractive force, so the $D3$-${\bar{D3}}$-brane inflation works nicely. This is the KKLMMT model~\cite{Kachru:2003sx}.

 The predictions of its inflaton potential $V(\phi)$ (where the inflaton $\phi$ measures the separation of the 2 branes) made years ago agree very well with the Cosmic Microwave Background (CMB) data obtained by the PLANCK collaboration~\cite{Planck:2016}. So it is worthwhile to examine the model more closely. 
 Since the full $D3$-${\bar{D3}}$-brane potential $U(T, \phi)$ of $\phi$ and a complex $T$ mode is determined, one can check its properties in some details and compare them with (present and future) cosmological observations. This is especially interesting since, besides  the production of cosmic superstrings, the possibility of a first order phase transition and the production of primordial black holes in the model after the inflationary epoch remains to be  explored. That is, the model can produce (or not produce) phenomenons detectable in cosmological observations and so can be critically tested. In fact, we consider its comparison with the existing CMB data as only a first test of the model, with other properties to be tested in the near future. In this paper, the full potential $U(T, \phi)$ is presented, followed by a brief discussion of its novel features. The actual phenomenological investigation is deferred to the future.
 	
	The inflation potential $U(T, \phi)$ of the $D3$-${\bar{D3}}$-brane model combines the results of Ref.\cite{Jones:2002sia,Sarangi:2003sg} (presented in Eq.(\ref{eq:SU}, \ref{Vpot0}), with its simplified version in Eq.(\ref{Vpot1}, \ref{ImV}, \ref{eq:realV})).
 The model is a function of 3 parameters, namely: \\
	 (1) the warped $D3$-brane tension $\tau_3$, which is determined by the Hubble parameter $H$ and the density perturbation first observed by COBE~\cite{COBE:1992syq}, and \\
	 (2) the string coupling $g_s \lesssim 1$. 
	 These 2 parameters are related by the scale $M_A$ of the warped throat. \\
	 (3) the curvature coupling $\zeta_{\phi} H^2\phi^2$ term is an issue encountered in all inflationary models. Here, PLANCK data~\cite{Planck:2016} implies that the parameter $|\zeta_{\phi}| < 0.01$. We shall simply set $\zeta_{\phi}=0$. \\
	 (The model also depends very weakly on the property of the warped throat in which the $D3$-${\bar{D3}}$-brane pair sits in.) 
	 
	 This is a hybrid inflation model~\cite{Linde:1993cn} in the sense that the slow-roll inflation happens when $T$ sits at its minimum $T=0$ and ends when $T$ becomes tachyonic and then rolls down the potential. It differs from the standard (and well studied) hybrid inflation models in a number of ways: \\
	$\bullet$ Based on a simple physical picture, the potential is well determined. 
 As a result of the $\phi^{-4}$ form of the attractive (gravitational + Ramond-Ramond (RR)) potential, it predicts a scalar power spectrum that is slightly red-tilt, in agreement with the CMB data from PLANCK~\cite{Planck:2016}.  \\
	$\bullet$ As the $D3$-brane approaches the ${\bar{D3}}$-brane ({\it i.e.}, $\phi$ decreases), $T$ becomes tachyonic and the $\phi^{-4}$ potential softens to a finite form (so the $\phi \to 0$ catastrophic divergence is absent), while $U(T, \phi)$ develops an imaginary component, indicating the decay (annihilation) of the $D3$-${\bar{D3}}$-brane pair. The corresponding decay width provides a much stronger damping than that from the Hubble parameter in the $\phi$ evolution. \\
	$\bullet$ Even as the $D3$-${\bar{D3}}$-branes move slowly towards each other during the inflationary epoch, $T$ can tunnel through a barrier and then roll to infinity. Nucleation bubbles will be created during this first order phase transition. Although this transition is unlikely to have a big cosmological impact in the simplest $D3$-${\bar{D3}}$-brane inflation model, such a first order phase transition phenomenon can leave behind observable signatures if there are multiple $D3$-${\bar{D3}}$-brane pairs in different warped throats.  The presence of multiple throats with different warp factors in the dynamical compactification of the 6 spatial dimensions is quite natural in the brane world scenario (e.g.,  {\it cf.}\cite{Polchinski:1998rq,Maartens:2003tw,Baumann:2014nda}).  
	
	Already, the model predicts the production of cosmic superstrings that form a network~\cite{Jones:2002cv,Sarangi:2002yt,Jones:2003da}, with properties rather different from field theory cosmic strings~\cite{Copeland:2003bj,Jackson:2004zg}. 
The production of primordial black holes (PBHs) within the framework of hybrid inflation is an interesting possibility~\cite{Randall:1995dj,Garcia-Bellido:1996mdl,Kawasaki:1997ju,Abolhasani:2010kr,Lyth:2011kj,Lyth:2012yp} and has been actively investigated (e.g., {\it cf.}\cite{Sasaki:2018dmp,Ozsoy:2023ryl}). Highly over-dense regions of inhomogeneities in the early Universe can directly undergo gravitational collapse to form black holes~\cite{Hawking:1971ei}. Physically, the collision and annihilation of a brane pair is an explosive process, reflected in the $(T, \phi)$ evolution as well as the presence of nucleation bubbles from a first order phase transition. Intuitively, large density fluctuations and gravitational waves are expected to be generated. Here we discuss briefly the general properties of the model and its novel features.
	
	The rest of the paper is organized as follows. Sec.~2 presents the $D3$-${\bar{D3}}$-brane action ${\cal S}(T, \phi)$ with the potential $U(T, \phi)$. This potential combines the result of Ref.\cite{Jones:2002sia,Sarangi:2003sg}. Sec.~3 gives a brief review of the properties during the inflationary epoch, where the predictions~\cite{Kachru:2003sx} are in good agreement with the PLANCK data~\cite{Planck:2016}. Sec.~4 give a brief description of some of the properties towards the end of the inflation epoch that have been extensively studied over the years. In Sec.~5, we discuss the role of the imaginary part ${\rm Im}\,U(T, \phi)$ in the dynamics. In Sec.~6, we discuss qualitatively the tunnelling aspects of the model, where nucleation bubbles from first order phase transition can come into play. Sec.~7 contains some remarks.

\section{The $D3$-${\bar{D3}}$-Brane Model}
	
With the string mass scale $M_S$ expressed in terms of the Regge slope $\alpha'$,
$M_S^2= 1/2 \pi \alpha'$, the 6-dim compactification volume ${\cal V}$ relates the string scale to the Newton's constant $G_N = 1/8 \pi M_P^2$, so $M_P=2.43 \times 10^{18}$ GeV and 
$$M_P^2 \sim M_S^8 {\cal V}$$
where $M_S \ll M_P$.
The $Dp$-brane tension is given by
\begin{equation}	
T_p=\frac{1}{(2\pi)^pg_s \alpha'^{(p+1)/2}}=\frac{M_S^{(p+1)}}{(2\pi)^{(p-1)/2} g_s}
\end{equation}
where we expect the string coupling to lie in the range $ 0.1 \le g_s < 1$. 

\subsection{The action ${\cal S}(T, \phi)$}

The $D3$-${\bar{D3}}$-brane inflation takes place in a warped throat, called the $A$-throat, while the standard model branes live in another throat, called the $SM$-throat.
 In the $A$-throat, the mass scale is warped to $M_A$, where 
 $$M_A \ll M_S \ll M_P$$
 so the warped $D3$-brane tension and the $D$-string tension in the A-throat are given by, respectively,
  \begin{equation}\label{eq:tension}
  \tau_3= \frac{M_A^4}{2\pi g_s}, \quad \quad \tau_1= \frac{M_A^2}{g_s},
\end{equation}
The $D3$-${\bar{D3}}$-brane potential has 2 parameters, namely $M_A$ and $g_s$. During the inflationary epoch, it is only $\tau_3$ that comes into play.

The $D{\bar{D}}$-brane action ${\cal S}(T)$ for tachyon condensation ({\it i.e.}, $T$ rolling down its potential) is derived in Ref.\cite{Kutasov:2000aq,Kraus:2000nj,Takayanagi:2000rz} in the Boundary Superstring Field Theory (BSFT)~\cite{Witten:1992qy,Witten:1992cr,Shatashvili:1993kk}. In Ref.\cite{Jones:2002sia}, via T-duality~\cite{Myers:1999ps}, the inflaton $\phi$ is incorporated to form the action ${\cal S}(T, \phi)$ for the $D3$-${\bar{D3}}$-brane pair system, with the Lagrangian density ${\cal L}(T, \phi)$,\footnote{The relative position of the 2 branes in 6 spatial dimensions involves more co-ordinates. They become (anti-)parallel after an exponential expansion via inflation; so they can be described by a single co-ordinate $\varphi$ as $D3$-${\bar{D3}}$-brane separation distance. We use $\phi =\sqrt{\tau_3}\varphi$ as the dynamical variable with a canonical kinetic term for slow-roll inflation.}
\begin{equation}\label{Upot}
{-\cal L}({\cal T}, \phi)= e^{-{\cal T}{\bar {\cal T}}/M_A^2}{\cal F}\left(\frac{\phi^2 {\cal T}{\bar {\cal T}}}{\pi \tau_3}\right){\cal F}\left(\frac{\partial_{\mu}{\cal T} \partial^{\mu}{\bar {\cal T}}}{\pi M_A^4} \right) V(\phi) + \frac{\partial_{\mu}\phi \partial^{\mu}\phi}{2}
\end{equation}
where $V(\phi)$ is the inflaton potential, and  $T$ is related to ${\cal T}$ by a rescaling.
All powers of the single derivative of $\cal T$ are contained inside ${\cal F}$~\cite{Kraus:2000nj}. The function ${\cal F}$ takes the form~\cite{Kutasov:2000aq}
\begin{equation}
	{\cal{F}}(x)= \frac{4^xx\Gamma(x)^2}{2 \Gamma(2x)}=\frac{\sqrt{\pi}\Gamma(1+x)}{\Gamma(\frac{1}{2} +x)}
\end{equation}
where
 \begin{align}\label{Fexpand}
{\cal{F}}(x) &= 1 +(2 \ln 2) x +\left[2 (\ln 2)^2 - \frac{\pi^2}{6} \right] x^2 + {\cal O}(x^3), \quad \quad 0 \le x \ll 1 \\
&= \sqrt{\pi x} \left[1 + \frac{1}{8x} + {\cal O}(x^{-2}) \right], \quad \quad x \gg 1
\end{align}
Expanding the function ${\cal F}(x)$ to leading order using Eq.(\ref{Fexpand}) and setting $V(\phi) \simeq 2 \tau_3$, we obtain the canonical kinetic term for $T$ (around $T=0$) :
\begin{equation}
\frac{\partial_{\mu}T \partial^{\mu}{\bar T}}{2}=\frac{(2 \ln 2)\partial_{\mu}{\cal T} \partial^{\mu}{\bar {\cal T}}}{\pi M_A^4}V(\phi) \simeq \frac{(4\ln 2)}{\pi ^2 g_s}\frac{\partial_{\mu} {\cal T} \partial^{\mu}{\bar {\cal T}}}{2}
\end{equation}
\begin{figure}[t]
 \begin{center}
  \includegraphics[width=4.5in]{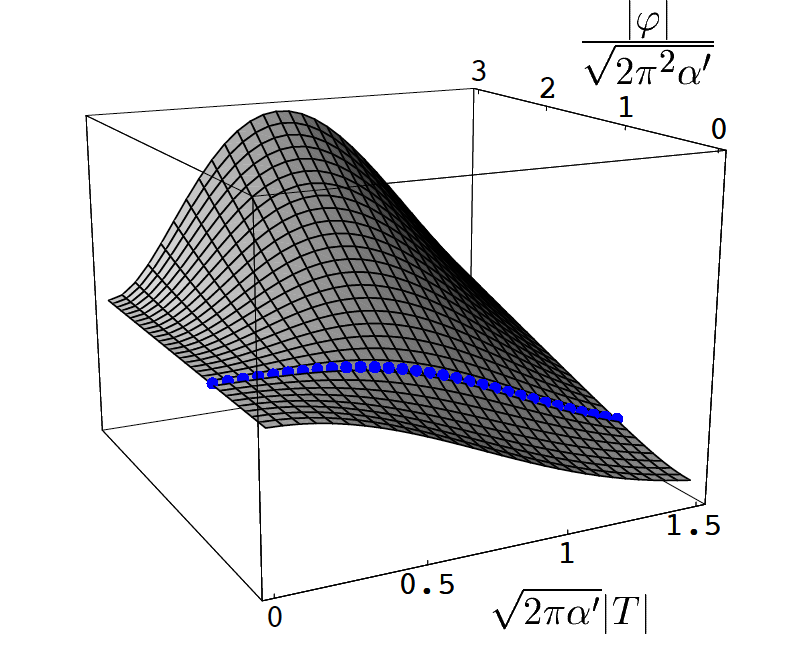}
  \caption{The $D3$-${\bar{D3}}$-brane potential $U(T, \varphi)$ where $\varphi= \phi/\sqrt{\tau_3}$ measures the separation distance between the 2 branes. The blue line indicates where $T$ becomes tachyonic as $\varphi$ decreases. Note that, even for a relatively large $\varphi$ where $T$ is classically stable at $T=0$, $U(T, \varphi) \to 0$ as $T \to \infty$. This figure is taken from Ref.\cite{Jones:2002sia}.}
 \label{figure}
   \end{center}
      \end{figure}
Ignoring higher powers of $\partial_{\mu}T \partial^{\mu}{\bar T}$, we now have
\begin{align}\label{eq:SU}
{\cal S}(T, \phi) &= -\int dx^4 a^3(t)\left(\frac{{\dot T} \dot{\bar T}}{2}  +\frac{{\dot \phi}^2}{2} 
+U(T,\phi) \right) \nonumber\\
U(T, \phi) &= \exp \left[-\frac{\pi^2 g_sT{\bar T}}{(4\ln 2)M_A^2}\right]{\cal{F}}\left(\frac{\pi g_s\phi^2 T{\bar T}}{(4 \ln 2)\tau_3}\right)V(\phi) 
\end{align}
where $a(t)$ is the cosmic scale factor and $\tau_3$ (\ref{eq:tension}) is a function of $M_A$ and $g_s$. The potential $U(T,\phi)$ is displayed in Figure 1 (taken from Ref.\cite{Jones:2002sia}). 

One may choose to keep all single $T$ derivative terms inside $\cal F$ in ${\cal S}(T, \phi)$(\ref{eq:SU}) and/or express the $\phi$ kinetic term in the Dirac-Born-Infeld form
\begin{equation}\label{DBI}
\tau_3\left(\sqrt{1 - {\dot\phi}^2/\tau_3} - 1 \right) \quad \to \quad {\dot \phi}^2/2
\end{equation}
which reduces to the canonical form in the slow-role approximation.

We note that the mass of $T$ is given by
\begin{equation}\label{tmass}
m_T^2(\phi)=\pi g_s\left(\frac{\phi^2}{\tau_3} -\frac{\pi} {(2 \ln 2)M_A^2}\right)V(\phi) \simeq
2 \pi g_s (\phi^2 - \phi_c^2)
\end{equation}
so $T$ becomes a tachyon when its mass square is negative, {\it i.e.}, when $\phi^2 < \phi_c^2$, where
\begin{equation}\label{Ttachyon}
\phi_c^2 = \frac{\pi \tau_3}{(2 \ln 2) M_A^2}= \frac{M_A^2}{(4 \ln 2) g_s}
\end{equation}

\subsection{The inflaton potential}

To complete, let us consider the potential $V(\phi)$ in $U(T, \phi)$ (\ref{eq:SU}) now. It is given by the $D3$-${\bar{D3}}$-brane tensions plus the interactive potential ${\hat V}(\phi)$, which is determined (especially for $\phi<\phi_c$~\cite{Sarangi:2003sg}). The term ${\hat V}(\phi)$ has a dual description (e.g., {\it cf.}\cite{Polchinski:1998rq}): it can be expressed as {\it either} an exchange of closed string modes between the 2 branes, where the lightest modes include the graviton and the RR field,
 {\it or} the one-loop correction to the open strings stretching between the 2 branes, where the lightest mode is $T$. Choosing the latter expression, we have
\begin{align}\label{Vpot0}
V(\phi)&= 2 \tau_3 + {\hat V}(\phi) \nonumber\\
{\hat V}(\phi) &= - \frac{v}{16} \left(\frac{\tau_3}{\phi_c^2}\right)^2 \int_0^{\infty} \frac{dt}{t^3}
\exp \left[- \frac{\pi t \phi^2}{\phi_c^2} \right] Z(t) \\
Z(t)&=e^{\pi t} \prod_{m=1}^{\infty} \frac{(1- e^{-2\pi(m-1/2)t})^8}{(1- e^{-2\pi mt})^8} =e^{\pi t}\ - 8 +36e^{-\pi t} + . . .\nonumber 
\end{align}
where the first term is the $D3$-${\bar{D3}}$-brane tension that drives inflation.
At large distances (large enough $\phi$), the above integral is dominated by small $t$, where $Z(t) \to 16 t^4$; so the potential reduces to
\begin{equation} \label{Vpot1}
V(\phi)=2 \tau_3 + {\hat V}(\phi)\simeq 2 \tau_3 \left(1 - v\frac{\tau_3}{2 \pi^2 \phi^4} \right), \quad \quad \phi>\phi_c 
\end{equation}
and the second term comes from the gravitational plus the RR attractive forces. The parameter $v \sim {\cal O}(1)$ measures the property of the throat. For the Klebanov-Strassler throat~\cite{Klebanov:2000hb}, $v=27/16$ (e.g., {\it cf.}\cite{Baumann:2014nda}), a value we shall adopt in our numerical estimates. 
Fortunately, the quantitative physics is insensitive to the value of $v$.

It is clear that $V(\phi)$ as given in Eq.(\ref{Vpot1}) breaks down as $\phi \to 0$. Fortunately, this disaster is actually absent in string theory. When $\phi < \phi_c$, ${\hat V}(\phi)$ (\ref{Vpot0}) diverges, because the lightest mode ({\it i.e.}, $e^{\pi t}$ in $Z(t)$) becomes tachyonic. This divergence is not an infinity of the theory, but rather a signal that ${\hat V}(\phi)$ becomes complex. After proper regularization by analytic continuation, ${\hat V}(\phi)$ develops an imaginary part as a function of $m_T^2$ (\ref{tmass}),
\begin{equation}\label{ImV}
{\rm{Im}}\, {\hat V}(\phi) = {|m_T^2|^2}/{32 \pi},
\end{equation}
indicating the annihilation (or decay) of the $D3$-${\bar{D3}}$-brane pair, while the real part ${\rm{Re}}\, {\hat V}(\phi)$ stays finite~\cite{Sarangi:2003sg}. It reaches a finite minimum ${\rm{Re}}\, {\hat V}(\phi_m)$ 
 at $\phi_m$ (where $\phi_c > \phi_m >0$) and $\phi$ becomes tachyonic ({\it i.e.}, $m_{\phi}^2 <0$) at $\phi =0$, a feature first observed in a similar situation~\cite{Garcia-Bellido:2001lbk}.  
Note that ${\rm{Re}}\, {\hat V}(\phi)$ is always much smaller than $2 \tau_3$, so it is a reasonable approximation to treat the Hubble value $H$ as a constant until $T$ starts to roll down. For our qualitative discussion in this paper, we need only the overall shape of ${\rm{Re}}\, {\hat V}(\phi)$ for $\phi < \phi_c$. We shall come back to its shape in Sec.~5.

\subsection{Discussions}

In this brane world scenario, all the energies released by the annihilation of the $D3$-${\bar{D3}}$-brane pair can only go to $D$-strings ({\it i.e.}, $D1$-branes) and the fundamental $F$-strings, since they are the only degrees of freedom left, {\it i.e.}, neither $D2$-branes nor $D0$-branes exist in Type IIB string theory. They are closed string modes since there are no branes (except $D1$-branes produced) around for the ends of open strings to end in.  The production of them can be seen more explicitly in the model :

(i) As an effective field theory, the model possesses a $U(1)$ symmetry: $T \to e^{i \theta}T$. Since string theory has no continuous global symmetry~\cite{Banks:1988yz}, such a symmetry is gauged. In fact, this gauged $U(1)$ property leads to the presence of topological vortex-like  (co-dimension 2) defects~\cite{Jones:2002sia}. These defects are to be identified as $D$-strings, which will be produced when $T$ rolls down the potential.

(ii) For large enough separation $\phi$, a $\phi^{-4}$ term in $V(\phi)$ is generated by the open string one-loop correction. In the dual closed string channel, this is due to the exchange of the graviton and the RR field, a classical effect. As $\phi$ decreases, the exchange of massive closed string modes becomes important. As $\phi$ further decreases, just as $T$ becomes tachyonic, $V(\phi)$ develops an imaginary part~\cite{Sarangi:2003sg}. By optical theorem, an ${\rm Im} \,V(\phi)$ indicates the opening of inelastic channels, {\it i.e.}, energies go to a spectrum of closed $F$-strings. That includes the graviton, some massive string modes, string loops of cosmological sizes as well as the light axion-like modes.

(iii) In an inflationary model, a general function $F(X)$ of $X=\partial_{\mu}\phi \partial ^{\mu} \phi$ is  known to generically cause problems for the reliability of its low energy expansion, with the exception of the DBI case (\ref{DBI}) where power counting shows control can be maintained as an expansion in powers of the sound speed. Here, since both $T=0$ and $\dot T=0$ during the inflationary epoch, both functions ${\cal F}$s in the model (\ref{Upot}) approach ${\cal F} \to 1$ and have no impact on the overall inflationary properties. They become important only towards the end of and after the inflationary epoch. On the other hand, if the inflationary epoch starts with a non-zero $T$, a small fluctuation of $T$ during the inflationary epoch should be examined for possible clues of additional features in the CMB.

To summarize, the $D3$-${\bar{D3}}$-brane pair potential is given by $U(T, \phi)$ (\ref{eq:SU}), inside where the inflaton potential $V(\phi)$ (\ref{Vpot0}) can be simplified to Eq.(\ref{Vpot1}) for $\phi \ge \phi_c$ (during inflation) and becomes complex for $\phi_c > \phi \ge 0$ (after inflation), with imaginary part ${\rm{Im}}\, {\hat V}(\phi)$ (\ref{ImV}), while the real part ${\rm{Re}}\, {\hat V}(\phi)$ is approximately given in Eq.(\ref{eq:realV}) in Sec.~5. 
This potential yields a hybrid inflation model, but with the following distinct features that we shall briefly explore : \\
(1) The potential $U(T,\phi)$ is well determined in terms of only 2 parameters, namely the scale $M_A$ and the brane tension $\tau_3$, which are related by the string coupling $g_s$. Other parameters (such as $v$) play very minor roles in confronting cosmological data; this allows a critical test of the model and check its properties.\\
(2) $U(T,\phi)$ develops a well determined imaginary part towards the end of the inflationary epoch, where ${\rm{Im}}\, {\hat V}(\phi)$ (\ref{ImV}) tends to damp the $\phi$ evolution. As this damping effect is much stronger than that from $H$, it can have a big impact on its subsequent evolution. \\
(3) As shown in Figure 1, $T$ can tunnel over the potential barrier and roll to infinity; this novel feature can play an important role when there are more than one $D3$-${\bar{D3}}$-brane pair.

\section{The Inflationary Epoch}

Let us briefly review the model's predictions for the CMB data and fix the value of the key parameter of the model, namely the $D3$-brane tension $\tau_3$ (\ref{eq:tension}).
We then explain why a curvature coupling term is absent in this $D3$-${\bar{D3}}$-brane model.

\subsection{The slow-roll inflationary epoch - a review}

When the $D3$-${\bar{D3}}$-brane separation is large enough, the $T$  mode is stable at $|T|=0$. So the potential (may be after some damped oscillation) reduces to $V(\phi)$ (\ref{Vpot1})
\begin{equation} \label{Vpot2}
U(T, \phi) \to V(\phi)=2 \tau_3 \left(1 - v\frac{\tau_3}{2 \pi^2 \phi^4} \right)= 2 \tau_3 \left(1 - \frac{C}{\phi^4} \right)
\end{equation}
where the Hubble parameter is
\begin{equation}\label{eq:H2}
H^2=V(\phi)/3M_P^2
\end{equation}
This model~\cite{Kachru:2003sx} has essentially only one parameter, namely $\tau_3$, which is determined by the measured density perturbation.
      
As the branes approach each other, $\phi$ decreases and inflation ends when $T$ becomes tachyonic, {\it i.e.}, when $\phi = \phi_c$ (\ref{Ttachyon}) (even though little more e-folds can happen after this, as we shall see). At $\phi = \phi_c$, the second term in $V(\phi_c)$ is still only a small fraction of the leading ($2 \tau_3$) term,
$$C/\phi_c^4 = \frac{27}{16 (2 \pi^4)} 2 \pi g_s (2 \ln 2)^2 \simeq 0.105 g_s $$
So during slow-roll inflation, the second term in $V(\phi)$ is even smaller. To a reasonable approximation, the slow-roll parameters are
\begin{align}
\epsilon &= \frac{M_P^2}{2}\left(\frac{V_{\phi}}{V}\right)^2 \simeq \frac{M_P^2}{2}\left(\frac{4C}{\phi^5}\right)^2\\
\eta &=M_P^2 V_{\phi \phi}/V \simeq -\frac{20 C M_P^2}{\phi^6}\\
\zeta &=M_P^4 \frac{V_{\phi} V_{\phi \phi \phi}}{V^2}
\end{align}
where $V_{\phi}= \frac{\partial V}{\partial \phi}$ and so on.
The density fluctuation $\delta_H$ (or the power spectral density $\sqrt{{\cal{P}}_k}$ measured at the pivot scale $k=aH=0.05 \,Mpc^{-1}$ where $a$ is the cosmic scale factor) as measured in the CMB by PLANCK~\cite{Planck:2016}, 
$${\cal{P}}(k)= \delta^2_H=(\delta \rho/\rho)^2 = 4.7 \times 10^{-9}$$
corresponds to about $N_e=50 - 60$ e-folds before slow-roll inflation ends at $\phi_f <\phi_c$. The number of e-folds $N$ is generated if we start at $\phi$ and end at $\phi_{f}$, 
$$N= \int_{\phi}^{\phi_f} \frac{H}{\dot{\phi}} d \phi =\frac{1}{M_P^2}\int_{\phi_f}^{\phi} \frac{V}{V'} d \phi$$
Slow-roll inflation ends when $\eta \simeq -1$ at $\phi_f= 24CM_P^2$,
\begin{equation}\label{phiN}
\phi^6 -\phi_f^6 = 24C M_P^2(N-1)
\end{equation}
so the value $\phi_i$ is given by
$$ \phi_i^6 \simeq 24C M_P^2N_e$$
 \begin{equation}
 \delta_H=\frac{1}{2 \pi}\left(\frac{H^2}{2 M_P^2 \epsilon}\right)^{1/2} \simeq \frac{1}{2 \pi} \left(\frac{\tau_3}{3 M_P^4 \epsilon}\right)^{1/2}\big|_{\phi_i}
 \end{equation}
 where $$\epsilon = 0.0022 \tau_3^{1/3}M_P^{-4/3}/N_e^{5/3}$$
 So we have
 $$\tau_3\simeq  
  \left(4.51 \times 10^{-5} M_P\right)^4$$
 or
 \begin{equation}\label{vM_A}
 M_A \simeq (2 \pi g_s)^{1/4} 4.5 \times 10^{-5}M_P \simeq (2 \pi g_s)^{1/4} \,1.08 \times 10^{14} \, {\rm GeV}
 \end{equation}
 and 
  \begin{equation}\label{vH}
 H\simeq 1.18 \times 10^{-9} M_P \simeq 2.6 \times 10^{-5} (2\pi g_s)^{-1/4} M_A
 \end{equation}
 We see that $\epsilon \ll \eta$.  The scalar power spectral index 
 \begin{equation}\label{eq:ns}
 n_s-1= \frac{d \ln {\cal{P}}_k}{d \ln k}= -6 \epsilon + 2\eta \simeq - \frac{5}{3 N_e} 
  \end{equation}
 so $$n_s = 0.967 - 0.972, \quad {\rm for} \quad N_e= 50 -60$$
 while PLANCK value is $n_s = 0.9677 \pm 0.0060$~\cite{Planck:2016}.
 The ratio of the tensor to scalar perturbation is negligibly small,
 $$r= 12.4 \epsilon \sim 10^{-10}$$
 and
  $$ \frac{dn_s}{d\ln k} = 16 \epsilon \eta -24 \epsilon^2 -2 \zeta \simeq  -\frac{5}{3 N_e^2} \sim 0.0006$$
  These values are in good agreement with the observed values by PLANCK~\cite{Planck:2016}.
  Note that the predicted values reflect the particular $-\phi^{-4}$ form of the potential.
    
  \subsection{Comment}
  
 In the brane world scenario, we assume all the moduli (K\"ahler, complex structure and the dilaton) in the 6-dimensional flux compactification have been stabilized~\cite{Giddings:2001yu,Kachru:2003aw}. The inflaton $\phi$ covers a small range of values ({\it i.e.}, this is a small field inflationary model) so its evolution should have little impact on the modulus values. On the other hand, the specific compactification will have an impact on $\phi$. As we turn on gravity (or open string one loop corrections)~\cite{DeWolfe:2002nn,Berg:2004ek,Baumann:2010sx}, the effective action will in general contain a curvature coupling,
\begin{equation}\label{eq:etaproblem}
V_{\cal R}(\phi) =\frac{\zeta_{\phi} }{12} {\cal R}\phi^2 = \zeta_{\phi} H^2 \phi^2
\end{equation}
where the value of the parameter $\zeta_{\phi}$ depends on the details of the warped geometry ({\it cf.}\cite{Baumann:2014nda} for a discussion on the presence of such a term within the string theory compactification framework). This term will lead to a disastrously large slow-roll parameter in slow-roll inflation. This is the $\eta$ problem present in many inflationary models. 

Including this term in the $D3$-${\bar{D3}}$-brane inflation~\cite{Firouzjahi:2005dh}, $n_s$ (\ref{eq:ns}) in the model is most sensitive to this term, which becomes
$$n_s = 1 + \frac{4 \zeta_{\phi}}{3} - \frac{5}{3 N_e}$$ 
Comparing to the CMB data~\cite{Planck:2016}, where $0.95 < n_s < 0.98$,  
we see that $$|\zeta_{\phi}| < 0.01$$
Phenomenologically, we can always ignore this term, {\it i.e.}, set $\zeta_{\phi}=0$. 
However, a term like $\zeta_T H^2T{\bar T}$ is entirely possible.

 \section{Towards the end of the inflationary epoch}
 
In this brane world scenario, all the energies released by the annihilation of the $D3$-${\bar{D3}}$-brane pair go to closed string modes: $D$-strings and the $F$-strings. 

 $\bullet$ Some energies are transferred from the closed string modes to particles ({\it i.e.}, microscopic open strings) in the stack of standard model 3-branes in the SM-throat. They reheat the observable universe to start the hot big bang~\cite{Barnaby:2004gg,Kofman:2005yz,Chialva:2005zy,Frey:2005jk,Chen:2006ni}. 
 
  $\bullet$ Some energies go to cosmological sized string loops. The $D$-string tension is determined (\ref{eq:tension}):
$$G\tau_1= \frac{1}{M_P^2}\left(\frac{\tau_3}{32 \pi g_s}\right)^{1/2} \simeq \frac{4}{\sqrt{g_s}} \times 10^{-10}$$ while the $F$-string tension $\mu$ is 
\begin{equation}\label{eq:mu}
G\mu = g_sG\tau_1 \simeq 4\sqrt{g_s} \times 10^{-10}
\end{equation} 
The precise amount of large string loops produced remains to be determined. Fortunately, if too little is produced, they evolve like $a^{-2}$; if too much is produced, they will chop themselves up quickly. As a result, they evolve to a scaling cosmic superstring network that scales like the radiation density during the radiation era and like matter density during the matter dominated era (e.g., {\it cf.}\cite{Vilenkin:2000}). 
The cosmic superstrings differ from field theory cosmic strings in 2 important aspects : 

	(1) A collection of $p$ $F$-strings and $q$ $D$-strings can form a $(p,q)$ bound state with tension $\mu_{(p,q)} \sim \mu \sqrt{p^2 + q^2/g_s^2}$, so there is a spectrum of string tensions, instead of a single tension \cite{Copeland:2003bj}; the bound on $p$ allows a set of strings to end at a bead~\cite{Firouzjahi:2006vp}.
	
	(2) While the intercommutation probability $P\simeq 1$ for field theory cosmic strings, it is  $10^{-3} \le P \le 1$ for $F$-strings and $0.1 \le P \le 1$ for $D$-strings~\cite{Jackson:2004zg}.
The amount of strings in the network must be increased by this factor in order for an increased number of collisions per unit length of string to offset the reduced $P$ in each collision.  
As a result, the string density scales like $1/P^{0.65} \to 1/P$~\cite{Sakellariadou:2004wq,Avgoustidis:2005nv},  where the $F$-string density will be enhanced most. Together with the fact that the number density of $\mu_{(p,q)}$ strings goes like~\cite{Tye:2005fn}
$$N_{(p,q)} \propto \mu_{(p,q)}^{-n}, \quad \quad 6 < n \le 10$$
the contribution of string bound states to the cosmic string density is suppressed. As a result, $F$-strings, with tension $G\mu \sim 4 g_s \times 10^{-10}$ should dominate the cosmic superstring network. The total fractional density is about $\Omega_{superstring} \sim 8 \pi {\cal G}G \mu$ where ${\cal G}\sim 10^3 - 10^4$.
	
Their decay to gravitational waves generates the stochastic gravitational wave background~\cite{Vilenkin:1981bx,Hogan:1984is,Blanco-Pillado:2013qja,Blanco-Pillado:2017rnf}, which might have already been observed in pulsar timing data~\cite{Ellis:2023tsl}. Their kinks and cusps generate bursts of gravitational waves~\cite{Damour:2001bk,Damour:2004kw,Suresh:2023hkz} (as well as bursts from the beads~\cite{Leblond:2009fq})  to be detected in the future. Another way of detection is via micro-lensing~\cite{Chernoff:2007pd}.
 
 $\bullet$ Some energies go to the light closed string modes, like those in the K\"ahler moduli, the complex structure moduli and/or the dilaton. Some can be ultra-light axion-like particles $a_i$, $i=1, 2, ...$. Because of its very light mass, $a_i$ is locked at its expectation value $\langle a_i \rangle$ away from its minimum during inflation. It will start rolling towards its minimum only when the Hubble parameter becomes smaller than its mass. Its rolling converts its vacuum energy to dark matter via the mis-alignment mechanism. A priori, there is no reason why $a_i$ takes the same $\langle a_i \rangle$ value everywhere. It is inflation that assures us that $\langle a_i \rangle$ has the same value everywhere.
 Some interesting possibilities with mass \\
 	(1) $m \sim 10^{-22}$ eV as fuzzy dark matter~\cite{Hu:2000ke,Schive:2014dra,Hui:2016ltb}; \\
	(2) $m \sim 10^{-20}$ eV as a second component of the fuzz dark matter to resolve some issues in the single component FDM~\cite{Luu:2018afg}; \\
	(3) $m \sim 10^{-30}$ eV to solve the $^7$Li problem in big bang nucleosynthesis and the Hubble tension~\cite{Fung:2021wbz}; \\
 	(4) $m < 10^{-30}$ eV where $m < H_{today}$, so its still frozen vacuum energy  contributes to the dark energy.

 \section{The Complex Potential}

The tachyon dynamics has been studied before~\cite{Sen:1998sm,Shiu:2002xp,Lambert:2003zr,Chen:2003xq,Sen:2004nf,Leblond:2006cc,Brandenberger:2007ca,Battefeld:2010rf}. In string theory, both $\phi$ and $T$ are open string modes in the $D3$-${\bar{D3}}$-brane system. Once they annihilate, $\phi$ and $T$ no longer exist as physical degrees of freedom.
In the effective field theory approximation, we see that $|T| \to \infty$ so both $\phi$ and $T$ drop out of any post-inflation dynamics. 

 As mentioned earlier, the $U(1)$ symmetry in  $T{\bar T}$ is gauged. Below, we shall simplify $T {\bar T} \to T^2$ where the radial $T$ is a real scalar mode.
This model contains similar structure to the hybrid inflation model \cite{Linde:1993cn,Randall:1995dj,Garcia-Bellido:1996mdl}, so we shall focus on the $D3$-${\bar{D3}}$-brane model's more novel features. 

To discuss the qualitative features, let us consider the $\phi^{-4}$ term in $V(\phi)$ (\ref{Vpot2}). Naively, it diverges badly as $\phi \to 0$. Fortunately, it does not drop to $\infty$ as $\phi \to 0$. Instead, as (real) $\phi$ decreases so $\phi$ becomes smaller than $\phi_c$ (\ref{Ttachyon}), $V(\phi)$  (or ${\hat V} (\phi)$) (\ref{Vpot2}) becomes complex, where ${\rm Im}\,{\hat V} (\phi)$ is given in Eq.(\ref{ImV}), while the real part smoothly approaches some finite value. Let us now find an approximate form to capture the shape of ${\rm{Re}}\, {\hat V}(\phi)$.

\subsection{${\rm{Re}}\, {\hat V}(\phi)$ for $\phi<\phi_c$}

Let ${\bar V}(\phi)={\rm{Re}}\, {\hat V}(\phi)$ be the real part of ${\hat V} (\phi)$ for $0 \le \phi \le \phi_c$.
As described in Ref.\cite{Sarangi:2003sg}, after $T$ becomes tachyonic as $\phi<\phi_c$, $\phi$ decreases towards $\phi_m >0$ where ${\bar V} (\phi)$ reaches its minimum ${\bar V}_{m}$ at $\phi=\phi_m$. As $\phi$ further decreases, ${\bar V} (\phi)$ increases and $\phi$ becomes tachyonic at $\phi=0$. Let us parameterize these qualitative features by a simple formula
\begin{equation}\label{interpolate}
{\bar V} (\phi) = a -b\phi^2 +c\phi^4, \quad \quad 0 \le \phi \le \phi_c
\end{equation}
where $b>0$ measures the tachyonic mass of $\phi$ at $\phi=0$. To match ${\bar V} (\phi)={\hat V} (\phi)$ at $\phi_c$, we have {\footnote{To match $V_{\phi}$ and $V_{\phi \phi}$ at $\phi_c$,  a formula more involved than ${\bar V} (\phi)$ (\ref{interpolate}) is required. For the sake of simplicity, this is not needed in our approximate analysis here.}}
$${\bar V}(\phi_c)=a - b \phi_c^2 + c \phi_c^4= {\hat V} (\phi_c) = - 2 \tau_3C/\phi_c^4=- v\tau_3^2/\pi^2\phi_c^4$$
Ref.\cite{Sarangi:2003sg} shows that $a={\bar V} (\phi=0)$ is comparable to ${\hat V} (\phi_c)$.
In a crude approximation to simplify the analysis, we simply set 
$$a=  {\hat V} (\phi_c) \quad \to \quad c = b/\phi_c^2$$
Now the minimum of ${\bar V} (\phi)$ is at
\begin{equation}\label{eq:minimum}
\phi_m^2=b/2c = \phi_c^2/2
\end{equation}
and 
$${\bar V} (\phi_m) = {\hat V} (\phi_c) - b \phi_c^2/4$$
Next, in another approximation, we set 
$${\bar V} (\phi_m) = 2{\hat V} (\phi_c)= -\frac{4 \tau_3 C}{\phi_c^4} \quad \quad |{\hat V} (\phi_c)|\simeq 0.0334 M_A^4$$  
So ${\bar V} (\phi)$ (\ref{interpolate}) is now given by, for $\phi_c \ge \phi \ge 0$, 
\begin{equation}\label{eq:realV}
{\bar V} (\phi)= c \left[\phi^4 -\phi_c^2\phi^2 -\frac{\phi_c^4}{4}  \right], \quad \quad   c=4|{\hat V}(\phi_c)|/\phi_c^4 \simeq 1.03 g_s^2
\end{equation}
Next we consider the imaginary part  ${\rm{Im}}\, {\hat V} (\phi)$ of ${\hat V} (\phi)$ (\ref{ImV}),
\begin{equation}
 {\rm{Im}}\, {\hat V} (\phi) =\frac{|m_T^2(\phi)|^2}{32 \pi}=\frac{\pi g_s^2}{8}\left(\phi^2 -\phi_c^2\right)^2
 \end{equation}
where the tachyon mass $m_T^2$ (\ref{tmass}) has been used. 
 As pointed out in Ref.\cite{Banks:1995ch}, the appearance of such an imaginary part indicates the opening up of inelastic channels and the violent annihilation can lead to the production of PBHs.

\section{The Complex Potential on the $T$ and the $\phi$ Evolution}

Now we examine the $\phi$ and $T$ evolutions, 
\begin{align}\label{eq:phi}
\ddot \phi + 3H\dot\phi + U_{\phi} &= 0 \\
 \ddot T + 3 H \dot T + U_{T} &=0 \label{eq:T}
\end{align}

\subsection{The $\phi$ Evolution}

Let us first consider the $\phi$ evolution (\ref{eq:phi}). Here we have
\begin{align}
{\rm Re} \,U_{\phi} &\simeq\exp \left[-\frac{\pi^2 g_sT^2}{(4\ln 2)M_A^2}\right] 2 \left(\pi g_s T^2 +2c\phi^2 -c\phi_c^2\right)\phi
\nonumber \\
{\rm Im}\, U_{\phi} &= \frac{g_s}{4}m^2_T(\phi) \phi =  \frac{\pi g_s^2}{2} \left(\phi^2 - \phi_c^2 \right) \phi 
\end{align}
Note that slow-roll inflation ends when $\eta_{\phi}=-1$ at $\phi=\phi_f$ before it reaches $\phi_c$, while the potential is still stable in the $T$ direction, {\it i.e.}, $\ddot T= \dot T = T=0$. So $\phi$ rolls down the potential quickly when it passes 
 $\phi_c$ until quantum diffusion $H/2 \pi$ drives $T$ away from $T=0$. 
Let us look at the situation when $\phi$ is rolling towards $\phi_m$ in the absence of quantum diffusion. For an order-of-magnitude 
qualitative picture of what is going on, consider the following simplified version for $\bar \phi=\phi-\phi_m$ around the minimum at $\phi_m$, where $m$ is the mass of $\phi$ at $\phi_m$,
\begin{align}\label{eq:barphi}
 \ddot {\bar \phi} &+ 3H\dot {\bar \phi} + (m^2 + i \gamma) {\bar \phi}=0 \nonumber \\
 m^2 &= 2c \phi_c^2 \simeq 0.74 g_s M_A^2    \\
 \gamma &= {\rm Im}\, U_{\phi}\big|_{\phi_m}\simeq - \frac{\pi g_s}{16 \ln 2}M_A^2 \simeq - 0.283 g_s M_A^2 \nonumber
 \end{align}
 Let 
 $$\lambda =m^2 -9H^2/4, \quad \quad \xi=\sqrt{\lambda^2 + \gamma^2}$$
 The solution to Eq.(\ref{eq:barphi}) is
 \begin{align}
  \bar \phi  \sim &\exp \left(-\left(\frac{3H}{2} + \left[\frac{\xi - \lambda}{2}\right]^{1/2}\right) t\right) \cos \left(\left[\frac{\xi + \lambda}{2}\right]^{1/2} t\right) \nonumber \\
  \sim & \exp \left(-\left[\frac{3H}{2} + \frac{|\gamma|}{2m}\right] t\right) \cos \left(mt \right) 
  \end{align}
  where the damping term plays the role of a decay width,
  $$ \Gamma_{\phi_m} = 2\left[\frac{\xi - \lambda}{2}\right]^{1/2} \simeq |\gamma|/m \sim 0.329 g_s^{1/2}M_A $$
  We may express the equations in a more familiar form, where ${\rm Im}\,U(T, \phi)$ is replaced by the decay width $\Gamma_{\phi}$ :
\begin{align}\label{wGamma}
\ddot {\bar \phi} + [3H +\Gamma_{\phi}]\dot{\bar \phi} + {\rm Re} \,U_{\phi} = 0
\end{align}
where
$$\Gamma_{\phi_m}/3H \approx 10^{4}g_s^{1/4}$$
  That is, the damping effect coming from the imaginary part of the potential dominates over the Hubble value. (We suspect a similar $\Gamma_{\phi}$ should be present in any hybrid inflation model.)
  Note that 
  $$0 \le \Gamma_{\phi}  \le 2\Gamma_{\phi_m} \quad {\rm as} \quad \phi_c \ge \phi \ge 0$$
 As $\phi$ rolls down from $\phi_f$ to $\phi_c$, $\gamma=0$, so its motion is damped only by $H$. Still, the slope is steep enough so this journey takes less than one e-fold. As $\phi$ moves further towards the minimum at $\phi_m$, $\Gamma_{\phi}$ grows, providing a much stronger damping than that coming from $H$. $\Gamma_{\phi}$ grows further as $\phi$ passes $\phi_m$. Here, $\phi$ oscillates with period $2 \pi/m$. In one cycle, its amplitude is suppressed by a factor of
  $$e^{-\Gamma_{\phi} (2 \pi/m)} \simeq e^{-5} \sim 0.007$$
so one expects $\phi$ to quickly settle down at $\phi_m$.

\subsection{The $T$ Evolution}

Next we consider the $T$ evolution (\ref{eq:T}). Since $m_T^2$ (\ref{tmass}) is independent of $T$, the derivative ${\rm Im}\,U_T =0$. We may extend $m_T^2$ to next order to obtain a non-trivial ${\rm Im}\,U_T$ and examine qualitatively its impact on the $T$ evolution.
To simplify the notations, we rewrite $U(T, \phi)$ (\ref{Upot}) with parameters $d$ and $f$,
$$U(T, \phi) \simeq 2 \tau_3 e^{-dT^2}{\cal F}\left(f\phi^2T^2 \right)$$
Expanding $U(T, \phi)$ up to 4$th$ power in $T$ and 
re-expressing ${\rm Re}\, U_{T}$ as
$${\rm Re}\, U_{T}= {\bar m}_T^2(T, \phi) T + \cdot \cdot \cdot $$
 the $T$ mass is qualitatively given by 
$$ {\bar m}_T^2(T, \phi) \simeq m_T^2(\phi) + 8 \tau_3T^2\left( d^2/2 - 2\ln 2 df\phi^2 + \left[2 (\ln 2)^2 - \frac{\pi^2}{6} \right]f^2\phi^4 \right)$$
where ${\bar m}_T^2(T=0, \phi)=m_T^2(\phi)$ (\ref{tmass}). 
Now,  ${\rm Im}\,U(T, \phi)$ extends to
$${\rm Im}\,U(T, \phi)= |{\bar m}_T^2|^2/32\pi$$
At $\phi=\phi_m$, we have its derivative
\begin{align}
{\rm Im}\,U(T, \phi_m)_T &\simeq  m_T^2(\phi_m)\frac{\tau_3 T}{\pi} \left(d^2/2 - \ln 2 df\phi_c^2 + \left[2 (\ln 2)^2 - \frac{\pi^2}{6} \right]\frac{f^2\phi_c^4}{4} \right) \nonumber\\
 &\sim m_T^2(\phi_m) T (- 0.056 g_s) \approx 0.13 M_A^2 T  =\sigma T
\end{align}
which is qualitatively valid for $(T/M_A)^2 < 0.4 g_s$.

We can get a qualitative feature of the solution of Eq.(\ref{eq:T}) by considering an approximate form ($m^2= -m_T^2 >0$)
 \begin{equation}\label{TGamma}
\ddot {T} + 3H\dot{T} +( - m^2 + i \sigma)T = 0 
\end{equation} 
In the absence of $H$ and $\sigma$, $T \propto e^{mt} \to \infty$ is the solution we are looking for. 
Following the same steps as above, we have (ignoring an oscillating factor)
$$T \approx \exp \left[(-\frac{3H}{2} + m + \frac{\sigma^2}{8 m^3})t\right]$$
Once $T$ starts rolling,  $T \to \infty$ quickly and $U(T, \phi) \to 0$ exponentially fast.
So the introduction of an ${\rm Im}\,U(T, \phi_m)_T$ has little impact on the qualitative behaviour on the $T$ evolution.

 The density fluctuation is given by
  $${\cal P}(k)\simeq \frac{1}{24 \pi} \frac{U}{M_P^4 \epsilon} \simeq \frac{1}{12 \pi}
\frac{U^3}{M_P^6 (U_T^2+U_{\phi}^2)}$$
Before $\phi$ reaches the minimum $\phi_m$, the quantum diffusion $\langle T \rangle \sim H/2\pi$ may start the $T$ rolling, which happens quickly. As a result, ${\cal P}(k)$ may never reach a very large value during the $(T, \phi)$ evolution.

  \section{Tunnelling Properties}

\subsection{Tunneling}

For fixed $\phi > \phi_c$, $T{\bar T}$ is classically stable at $T=0$. However, as $T{\bar T}$ increases for fixed $\phi$, $U(T, \phi)$ rises to a peak and then comes down so  $U(T, \phi) \to 0$ as $T{\bar T} \to \infty$ (see Figure 1). Although $T{\bar T}=0$ is classically stable, it can tunnel over the barrier and roll to $\infty$. Although the branes are still separated by distance $\phi>\phi_c$, tunnelling means a nucleation bubble forms at a point where the branes fluctuate enough so they collide and begin to annihilate. The growth of the bubble is slower than the inflation speed~\cite{Guth:1982pn}, so the complete brane annihilation happens only after $\phi \le \phi_c$. 

The probability of tunnelling is given by $ P(\phi)  \simeq K(\phi) e^{-S(\phi)}$, where $S$ is the Euclidean action of the instanton (or the ``bounce" solution). $S(\varphi)$ for $Dp-{\bar{Dp}}$-brane pair system has been estimated \cite{Jones:2002sia}, for $\varphi \ge \varphi_c$,
\begin{equation}\label{Nick}
S(\varphi) \simeq  4 \pi c_1c_2^{p+1} \left[\frac{p^p}{(p+1)g_s} \frac{2 \pi^{(p+1)/2}}{\Gamma(\frac{p+1}{2})}\right]\left(\frac{\varphi - \varphi_c}{\sqrt{\alpha'}}\right)^{(p+1)/2}
\end{equation}
where $c_1\simeq 1.5$ and $c_2 \simeq 0.29$. Note that the exponential tunnelling suppression vanishes when the barrier disappears at $\varphi=\varphi_c$, as expected.
The thin wall approximation~\cite{Coleman:1977} turns out to be  good when $\varphi$ is large. 
Applying it to our $D3$-${\bar{D3}}$-brane pair case ($\phi = \sqrt{\tau_3}\varphi$) and using Eq.(\ref{phiN}), where $\phi-\phi_c \simeq 24C M_P^2N/32\phi_c^5$ for $\phi$ close to $\phi_c$, 
\begin{equation} \label{tunnel1}
S (\phi) \simeq 10^{3} \frac{\left(\phi-\phi_c\right)^2}{M_A^2} \sim \left(\frac{M_P}{M_A}\right)^4 N^2
\end{equation}
where $N$ is the number of e-folds before $\phi$ reaches $\phi_c$, {\it i.e.}, $\phi-\phi_c \propto N$.
Let the probability of tunnelling per unit volume at $\phi$ during the inflationary epoch goes like
$P(\phi) \sim M_A^4 \exp (-S)$ in a volume $e^{3(N_e-N)}/M_A^3$, we have the probability per unit time
$${\cal P}(N) \sim M_A \exp \left[3(N_e-N) - (M_P/M_A)^4 N^2\right]$$
For $N\sim N_e$, the second term severely suppresses the tunnelling. As $N \to 0$, the  exponential increase in volume from inflation is not able to overcome the exponential suppression until $N$ is very small. 
However, the time $\delta t \sim N/H$ available for tunnelling $T$ starts rolling is too short to make much difference. 

The above estimate does not include gravitational effect. Turning on gravity leads us to the CDL tunnelling~\cite{Coleman:1980aw}. 
While the $H/2\pi$ fluctuation provides some enhancement to the tunnelling rate, any radiation inside the bubble will suppress the tunnelling process~\cite{Huang:2008jr}. Both the $\phi$ and the $T$ evolutions may further complicate the dynamics of this phase transition.

\subsection{Double $D3$-${\bar{D3}}$-brane pairs}

Since the compactified manifold presumably has a number of throats with different warp factors, one can easily imagine 
$D3$-branes and ${\bar{D3}}$-branes move towards the throats and fall in.  In any throat with
at least one $D3$-brane and one ${\bar{D3}}$-brane, they will approach each other and eventually annihilate until only the remaining $D3$-branes or ${\bar{D3}}$-branes are left behind. Suppose, not only a $D3$-${\bar{D3}}$-brane pair is present in the $A$-throat, but another $D3$-${\bar{D3}}$-brane pair is present in another throat, say the $B$-throat. This can be  an example of double inflation~\cite{Silk:1986vc}, with the possibility of PBH formation (e.g., {\it cf.}\cite{Sasaki:2018dmp,Ozsoy:2023ryl}).  

Most of the formulae for the pair in the $A$-throat applies to the $B$-throat pair, with $M_A$ and $\tau_3$ replaced by $M_B$ and $\tau_{3B}$, keeping the same string coupling $g_s$. 
Not to interfere with the properties of the $D3$-${\bar{D3}}$-brane inflation in the $A$-throat, let the $B$-throat warping be stronger, so 
$$M_A \gg M_B, \quad \quad V_A(\phi_A) \gg V_B(\phi_B)$$
In this case, the separation $\phi_B$ from its critical value $\phi_{Bc}$ will be frozen until $H \to \sqrt{V_B(\phi_B)/3M_P^2}$. Now there is more time for $T_B$ tunnelling over the $T_B$ barrier (see Figure 1) so this first order phase transition is more likely to happen.
Here, the Euclidean action for $T_B$ tunnelling through its barrier in the $B$-throat is given by
$$S_B \simeq 10^{3} \frac{\left(\phi_B-\phi_{Bc}\right)^2}{M_B^2}, \quad \phi_{Bc}^2=\frac{M_B^2}{(4\ln 2) g_s}$$
Let us consider the general case and then the special case where the $B$-throat is identified with the standard model $SM$-throat.

We start with the scenario when both $D3$-${\bar{D3}}$-brane pairs have relatively large separations, {\it i.e.}, $\phi_A >\phi_{Ac}$ and $\phi_B > \phi_{Bc}$. Although the $T_B$ barrier height is a function of $\phi_B$, it is qualitatively of order $M_B^4$. 
Hawking-Moss tunnelling~\cite{Hawking:1981fz}  goes like, order-of-magnitude-wise,
\begin{equation}
S_B \approx 24 \pi^2M_P^4\left(\frac{1}{M_A^4} - \frac{1}{M_A^4 +M_B^4}\right) \sim 24 \pi^2M_P^4 \frac{M_B^4}{M_A^8}
\end{equation}
With $M_A \simeq 5\times 10^{-5}M_P$ (\ref{vM_A}), we see that, for 
$$M_B \le 10^{-10}M_P \quad \rightarrow \quad S_B \sim 0$$
So the tunnelling rate $P\sim e^{-S_B}$ is not exponentially suppressed. In fact, bubbles may be profusely produced.
{\it i.e.},
During inflation in the $A$-throat, $H\sim 10^{-9}M_P$ (\ref{vH}) so $H > M_B$. In this case, $\dot{\phi}_B \simeq V_{B \phi_B}/3H$ is negligibly small so $\phi_B$ is essentially frozen, while  
the ``B" nucleation bubbles are produced quickly. Because of the rapid ({\it i.e.}, exponential) expansion of the universe, these bubbles do not collide. After the ``A" throat inflationary epoch, the universe enters a radiation-dominated era, when $\phi_B$ remains frozen. There are 2 possibilities:

(1) If $S_B$ remains small enough for a long enough time period, the production of ``B" nucleation bubbles remains to be efficient. Their rapid growth can lead to bubble collisions in this radiation-dominated era and complete the ``B" throat $D3$-${\bar{D3}}$-brane annihilation without going through another inflationary epoch.

(2) If $S_B$ is big enough so the bubble production is now suppressed. As $H$ approaches $H^2 \sim V_B(\phi_B)/3M^2$, the universe moves from a radiation-dominated epoch to a slow-roll $\phi_B$ inflationary epoch. Here, the bubbles already produced are too far apart to collide. Eventually, $\phi_B$ rolls pass $\phi_{Bc}$ so the $B$-throat brane pair collides and annihilates. As the $T_B$ potential barrier disappears, the bubbles already produced may simply fizzle out before they have any chance to collide. 

Depending on the details, it remains to be studied whether this phase transition and the formation of nucleation bubbles can yield observable gravitational radiation and lead to big enough density perturbations for the PBH formation, and/or other cosmological detectable signatures.

\subsection{$D3$ + Stack of ${\bar{D3}}$-branes in the Standard Model throat}

In general, we can consider a system of $n$ $D3$-branes and $m$ ${\bar{D3}}$-branes.  Such a system is quite natural in the brane world scenario. One possibility is to identify the $B$-throat with the standard model $SM$-throat.
Suppose we live in a stack of 5 ${\bar{D3}}$-branes today, which is big enough to accommodate the standard $SU(3) \times SU(2) \times U_Y(1)$ model. 
In early universe, there is a 
$D3$-brane moving towards a stack of 6 ${\bar{D3}}$-branes inside this throat. After colliding with the stack of ${\bar{D3}}$-branes, the $D3$-brane annihilates with one of the ${\bar{D3}}$-branes, leaving behind the stack of 5 ${\bar{D3}}$-branes we live in today. The energy released can go directly to the open string modes inside the stack, {\it i.e.}, SM particles that reheat our universe. In this case, the reheating of our universe is much more efficient than transferring energy from the $A$-throat to the $SM$ throat. Note that the electroweak phase transition can remain second order (or weakly first order). For example, the effective potential for the electroweak Higgs doublet $\Phi$ can take the standard form at temperature ${\cal T}$,
$V(\Phi) = \left(\kappa {\cal T}^2 - m^2 \right){\bar \Phi}\Phi + \lambda \left(\Phi^{\dag}\Phi \right)^2$.  

One possibility is suggested by the investigation in Ref.\cite{Jones:2003ae}. Consider $(5+n)$ ${\bar{D3}}$-branes parallel to $n$
$D3$-branes. The mode $T$ transforms in the bi-fundamental $(5+n, {\bar n})$ representation of $U(5+n)\times U(n)$. Since QCD plays a spectator role only, we may ignore it. With $T$ now in the $(2+n, {\bar n})$ representation of 
$$U(2+n)\times U(n) \to U(2) \times U(n) \times U(n)$$ where the $T$ in the $(1, n, {\bar n})$ representation rolls down the potential ({\it i.e.}, $T \to \infty$) as $n$ ${\bar{D3}}$-branes annihilate with the $n$ $D3$-branes. Any mode transforming non-trivially under either $U(n)$ will also drop out of the spectrum. The left-over 2 ${\bar{D3}}$-branes contain the electroweak $SU(2)$, where the $T$ in the $(2, 1,1)$ representation may be identified with the electroweak Higgs doublet $\Phi$. It will be interesting to see how the first order phase transition impacts on the electroweak second order phase transition.

\section{Remarks}

The recent discovery of the stochastic gravitational wave background (SGWB)~\cite{NANOGrav:2023gor,NANOGrav:2023hde,Xu:2023wog,EPTA:2023sfo,Zic:2023gta} has led to an intense investigation of its origin. Some of its possible sources are (i) cosmic strings (e.g.,\cite{Blanco-Pillado:2017rnf,Ellis:2023tsl}), (ii) massive black hole binaries (e.g., {\it cf.}\cite{NANOGrav:2023gor,Ellis:2023oxs}) and (iii) first order phase transitions (e.g.,\cite{Witten:1984rs,Hogan:1986qda,Turner:1990rc,Kamionkowski:1993fg,Chialva:2010jt,Hindmarsh:2015qta,Jiang:2015qor,An:2020fff}). It is interesting that the $D3$-${\bar{D3}}$-brane inflation model offers cosmic superstrings as well as first order phase transitions as possible sources of SGWB, while the production, the growth and/or the collision of nucleation bubbles may lead to large density perturbations resulting in the PBH production. It is important to explore the model in further details. As a start, the $D3$-${\bar{D3}}$-brane potential $U(T, \phi)$ is presented in this paper. We see that $U(T, \phi)$ has distinct novel features signifying its string theory properties. It is also well determined, allowing detailed analyses to test the model.  With a clear physical picture behind it, one can modify or extend the model in ways where the dynamics is still under control. 
Besides extending the single $D3$-brane (and/or the single ${\bar{D3}}$-brane) to a stack, or extending the model to multiple $D3$-${\bar{D3}}$-brane pairs in different warped throats, one may also consider other effects : \\
$\bullet$ Revert to a Dirac-Born-Infeld action~\cite{Alishahiha:2004eh,Shandera:2006ax}; the DBI action can play an important role towards the end of inflationary epoch when $\phi$ exits the slow-roll phase. \\
$\bullet$ Although CMB data from PLANCK implies that the term $H^2\phi^2$ is either absent or negligibly small, one may include a $\zeta_{T}H^2T {\bar T}$ term into $U(T, \phi)$. Among other effects, its presence tends to shift the value of $\phi_c$.\\
$\bullet$ Other (quantum and stringy) corrections to ${\cal S}(T, \phi)$.\\
$\bullet$ The interplay between cosmic superstrings, PBHs and nucleation bubbles from a first order phase transition can be important in the SGWB production.\\
The success of the model's predictions from the inflationary epoch (in confronting the CMB data) suggests that its life after the slow-roll inflationary epoch should be equally rich and exciting.

\begin{acknowledgments}
		This work is based on early collaborations with Sash Sarangi and Nick Jones, to whom I am grateful. I thank Xingang Chen, Hassan Firouzjahi, Tao Liu and especially Liam McAllister for valuable comments.
		
\end{acknowledgments}


\begin{thebibliography}{1999}
	
	\bibitem{Guth:1980zm}
	A.~H.~Guth,
	``The Inflationary Universe: A Possible Solution to the Horizon and Flatness Problems",
	Phys. Rev. D23, 347 (1981).
	
	\bibitem{Linde:1981mu}
	A.~Linde,
	``A New Inflationary Universe Scenario: A Possible Solution of the Horizon, Flatness, Homogeneity, Isotropy and Primordial Monopole Problems",
	Phys. Lett. B108, 389 (1982).
	
	\bibitem{Dvali:1998pa} 
G.~R.~Dvali and S.-H.~H.~Tye, 
``Brane inflation", 
Phys.\ Lett.\ B{450}, 72 (1999)  [hep-ph/9812483].

\bibitem{Dvali:2001fw} 
  G.~R.~Dvali, Q.~Shafi and S.~Solganik,
  ``D-brane inflation",
  arXiv: hep-th/0105203.
  
\bibitem{Burgess:2001fx} 
  C.~P.~Burgess, M.~Majumdar, D.~Nolte, F.~Quevedo, G.~Rajesh and R.~J.~Zhang,
  ``The Inflationary brane anti-brane universe",
  JHEP {0107}, 047 (2001)
  [hep-th/0105204].
  
   \bibitem{Giddings:2001yu}
    S.~B.~Giddings, S.~Kachru and J.~Polchinski,
    ``Hierarchies from fluxes in string compactifications",
    Phys. Rev. D66, 106006 (2002) [hep-th/0105097].

    \bibitem{Kachru:2003sx}
S.~Kachru, R.~Kallosh, A.~Linde, J.~Maldacena, L.~McAllister and S.~P.~Trivedi, 
``Towards inflation in string theory",
JCAP {0310}, 013  (2003) [hep-th/0308055].

  \bibitem{Planck:2016} 
  PLANCK Collaboration: P. A. R. Ade et al., Astron. Astrophys. 594, A20 (2016) [1502.02114].
  
	 \bibitem{Jones:2002sia} 
  N.~T.~Jones and S.-H.~H.~Tye,
  ``An Improved brane anti-brane action from boundary superstring field theory and multivortex solutions",
  JHEP {0301}, 012 (2003) [hep-th/0211180].
  
 \bibitem{Sarangi:2003sg} 
  S.~Sarangi and S.-H.~H.~Tye,
  ``Interbrane potential and the decay of a nonBPS D-brane to closed strings",
  Phys.\ Lett.\ B{573}, 181 (2003)  [hep-th/0307078].
  
\bibitem{COBE:1992syq}
    G.~F.~Smoot and others,
    ``Structure in the COBE differential microwave radiometer first year maps",
    Astrophys. J. Lett. 396, L1 (1992).

 \bibitem{Linde:1993cn}
   A.~D.~Linde,
    ``Hybrid inflation",
        Phys. Rev. D49, 748 (1994) [astro-ph/9307002].
        
        \bibitem{Polchinski:1998rq}
    J.~Polchinski,
    \underline{String theory}
    Cambridge University Press (2007).
    
    \bibitem{Maartens:2003tw}
   R.~Maartens, 
    ``Brane world gravity",
   Living Rev. Rel. 7, 7 (2004) [gr-qc/0312059].

     \bibitem{Baumann:2014nda}
    D.~Baumann and L.~McAllister,
    \underline{Inflation and String Theory}
    Cambridge University Press (2015) [1404.2601].
 
\bibitem{Jones:2002cv} 
  N.~T.~Jones, H.~Stoica and S.-H.~H.~Tye,
  ``Brane interaction as the origin of inflation",
  JHEP {0207}, 051 (2002) [hep-th/0203163].
  
\bibitem{Sarangi:2002yt} 
  S.~Sarangi and S.-H.~H.~Tye,
  ``Cosmic string production towards the end of brane inflation",
  Phys.\ Lett.\ B{536}, 185 (2002) 
  [hep-th/0204074].
  
\bibitem{Jones:2003da} 
  N.~T.~Jones, H.~Stoica and S.-H.~H.~Tye,
  ``The Production, spectrum and evolution of cosmic strings in brane inflation",
  Phys.\ Lett.\ B{563}, 6 (2003)
  [hep-th/0303269].
  
  \bibitem{Copeland:2003bj}
E.~J.~Copeland, R.~C.~Myers and J.~Polchinski,
``Cosmic F- and D-strings",
JHEP {0406}, 013 (2004) [hep-th/0312067].

   \bibitem{Jackson:2004zg}
M.~G.~Jackson, N.~T.~Jones and J.~Polchinski,
``Collisions of cosmic F- and D-strings",
JHEP {0510}, 013 (2005) [hep-th/0405229].

        \bibitem{Randall:1995dj}
   L.~Randall, M.~Soljacic and A.~H.~Guth,
    ``Supernatural inflation: Inflation from supersymmetry with no (very) small parameters",
    Nucl. Phys. B472, 377 (1996) [hep-ph/9512439].

\bibitem{Garcia-Bellido:1996mdl}
    J.~Garcia-Bellido, A.~D.~Linde and D.~Wands,
   ``Density perturbations and black hole formation in hybrid inflation",
   Phys. Rev. D54, 6040 (1996) [astro-ph/9605094].

\bibitem{Kawasaki:1997ju}
   M.~Kawasaki, N.~Sugiyama and T.~T.~Yanagida,
    ``Primordial black hole formation in a double inflation model in supergravity",
    Phys. Rev. D57, 6050 (1998) [hep-ph/9710259].
    
  \bibitem{Abolhasani:2010kr}
    A.~A.~Abolhasani and H.~Firouzjahi,
    ``No Large Scale Curvature Perturbations during Waterfall of Hybrid Inflation",
    Phys. Rev. D83, 063513 (2011) [1005.2934].

  \bibitem{Lyth:2011kj}
    D.~H.~Lyth,
    ``Primordial black hole formation and hybrid inflation",
    arXiv:1107.1681 [astro-ph.CO].

 \bibitem{Lyth:2012yp}
    D.~H.~Lyth,
    ``The hybrid inflation waterfall and the primordial curvature perturbation",
    JCAP05, 022 ( 2012) [1201.4312].

\bibitem{Sasaki:2018dmp}
   M.~Sasaki, T.~Suyama,  T.~Tanaka, and S.~Yokoyama, 
    ``Primordial black holes\textemdash{} - perspectives in gravitational wave astronomy",
 Class. Quant. Grav. 35, 6, 063001 (2018). [1801.05235].
 
 \bibitem{Ozsoy:2023ryl}
    O.~\"Ozsoy and G.~Tasinato,
    ``Inflation and Primordial Black Holes",
       Universe, 9, 5, 203 (2023) [2301.03600].

\bibitem{Hawking:1971ei}
    S.~Hawking,
    ``Gravitationally collapsed objects of very low mass",
    Mon. Not. Roy. Astron. Soc. 152, 75 (1971).

\bibitem{Kutasov:2000aq}
    D.~Kutasov, M.~Marino and G.~Moore,
    ``Remarks on tachyon condensation in superstring field theory",
    arXiv:hep-th/0010108.

   \bibitem{Kraus:2000nj}
    P.~Kraus and F.~Larsen,
    ``Boundary string field theory of the D anti-D system",
    Phys. Rev. D63, 106004 (2001) [hep-th/0012198].

  \bibitem{Takayanagi:2000rz}
    T.~Takayanagi, S.~Terashima and T.~Uesugi,
    ``Brane - anti-brane action from boundary string field theory",
     JHEP 03, 019 (2001) [hep-th/0012210].

\bibitem{Witten:1992qy}
    E.~Witten,
    ``On background independent open string field theory",
    Phys. Rev. D46, 5467 (1992) [hep-th/9208027].
    
   \bibitem{Witten:1992cr}
    E.~Witten,
   ``Some computations in background independent off-shell string theory",
   Phys. Rev. D47, 3405 (1993) [hep-th/9210065].

\bibitem{Shatashvili:1993kk}
    S.~L.~Shatashvili,
   ``Comment on the background independent open string theory",
    Phys. Lett. B311, 83 (1993) [hep-th/9303143].

 \bibitem{Myers:1999ps} 
  R.~C.~Myers,
  ``Dielectric branes'',
  JHEP 9912, 022 (1999) [hep-th/9910053].
   
\bibitem{Klebanov:2000hb}
I.~R.~Klebanov and M.~J.~Strassler,
``Supergravity and a confining gauge theory: Duality cascades and
chi-SB-resolution of naked singularities", 
JHEP {0008}, 052 (2000) [hep-th/0007191].

 \bibitem{Garcia-Bellido:2001lbk}
    J.~Garcia-Bellido, R.~Rabadan and F.~Zamora,
    ``Inflationary scenarios from branes at angles",
    JHEP 01, 036 (2002) [hep-th/0112147].

\bibitem{Banks:1988yz}
    T.~Banks and L.~J.~Dixon,
    ``Constraints on String Vacua with Space-Time Supersymmetry",
    Nucl. Phys. B307, 93 (1988).
    
\bibitem{Kachru:2003aw}
    S.~Kachru, R.~Kallosh, A.~D.~Linde and S.~P.~~Trivedi,
    ``De Sitter vacua in string theory",
      Phys. Rev. D68, 046005 (2003) [hep-th/0301240]

\bibitem{DeWolfe:2002nn}
   O.~DeWolfe and S.~B.~Giddings,
    ``Scales and hierarchies in warped compactifications and brane worlds",
     Phys. Rev. D67, 066008 (2003) [hep-th/0208123].
    
   \bibitem{Berg:2004ek}
   M.~Berg, M.~Haack and B.~Kors,
    ``Loop corrections to volume moduli and inflation in string theory",
    Phys. Rev. D71, 026005 (2005) [hep-th/0404087].

   \bibitem{Baumann:2010sx}
   D.~Baumann, A.~Dymarsky, S.~Kachru, I.~R.~Klebanov and L~McAllister,
   ``D3-brane Potentials from Fluxes in AdS/CFT",
    JHEP 06, 072 (2010) [1001.5028].
    
\bibitem{Firouzjahi:2005dh}
    H.~Firouzjahi and S.-H.~H.~Tye,
    ``Brane inflation and cosmic string tension in superstring theory",
    JCAP 03, 009 (2005) [hep-th/0501099]
 
\bibitem{Barnaby:2004gg}
    N.~Barnaby, C.~P.~Burgess and J.~M.~Cline,
    ``Warped reheating in brane-antibrane inflation",
    JCAP 04, 007 (2005) [hep-th/0412040].

\bibitem{Kofman:2005yz}
   L.~Kofman and P.~Yi,
   ``Reheating the universe after string theory inflation",
    Phys. Rev. D72, 106001 (2005) [hep-th/0507257].

\bibitem{Chialva:2005zy}
    D.~Chialva, G.~Shiu and B.~Underwood,
   ``Warped reheating in multi-throat brane inflation",
    JHEP 01, 014 (2006) [hep-th/0508229].

\bibitem{Frey:2005jk}
    A.~R.~Frey, A.~Mazumdar and R.~C.~Myers,
    ``Stringy effects during inflation and reheating",
     Phys. Rev. D73, 026003 (2006) [hep-th/0508139].

\bibitem{Chen:2006ni} 
  X.~Chen and S.-H.~H.~Tye,
  ``Heating in brane inflation and hidden dark matter",
  JCAP {0606}, 011 (2006)  [hep-th/0602136].
    
	\bibitem{Vilenkin:2000}
A.~Villenkin and E.~P.~S.~Shellard, \underline{Cosmic strings
and other topological defects}, Cambridge University Press (2000).  

\bibitem{Firouzjahi:2006vp}
   H.~Firouzjahi, L.~Leblond and S.-H.~H.~Tye,
    ``The (p,q) string tension in a warped deformed conifold",
   JHEP 05, 047 (2006) [hep-th/0603161].

\bibitem{Sakellariadou:2004wq} 
  M.~Sakellariadou,
  ``A Note on the evolution of cosmic string/superstring networks",
  JCAP {0504}, 003 (2005)
  [hep-th/0410234].
  
   \bibitem{Avgoustidis:2005nv}
    A.~Avgoustidis and E.~P.~S.~Shellard",
    ``Effect of reconnection probability on cosmic (super)string network density",
    Phys. Rev. D73, 041301 (2006) [astro-ph/0512582].

\bibitem{Tye:2005fn} 
  S.-H.~H.~Tye, I.~Wasserman and M.~Wyman,
  ``Scaling of multi-tension cosmic superstring networks",
  Phys. Rev. D{71}, 103508 (2005)
  [Erratum-ibid. D{71}, 129906 (2005)]  [astro-ph/0503506].
  
  \bibitem{Vilenkin:1981bx}
    A.~Vilenkin,
    ``Gravitational radiation from cosmic strings",
    Phys. Lett. B107, 47 (1981).

\bibitem{Hogan:1984is}
    C.J.~Hogan and M.~J.~Rees,
       ``Gravitational interactions of cosmic strings",
     Nature 311, 109 (1984).

\bibitem{Blanco-Pillado:2013qja}
   J.J.~Blanco-Pillado, K.D.~Olum and B.~Shlaer,
    ``The number of cosmic string loops",
    Phys. Rev. D89, 2, 023512 (2014) [1309.6637].

   \bibitem{Blanco-Pillado:2017rnf}
    J.~J.~Blanco-Pillado, K.~D. Olum and X. Siemens,
    ``New limits on cosmic strings from gravitational wave observation",
    Phys. Lett. B778, 392 (2018) [1709.02434].

\bibitem{Ellis:2023tsl}
    J.~Ellis, John M.~Lewicki, C.~ Lin and V.~Vaskonen,
    ``Cosmic Superstrings Revisited in Light of NANOGrav 15-Year Data",
    arXiv: 2306.17147 [astro-ph.CO].

\bibitem{Damour:2001bk} 
  T.~Damour and A.~Vilenkin,
  ``Gravitational wave bursts from cusps and kinks on cosmic strings",
  Phys. Rev. D {64}, 064008 (2001)
  [gr-qc/0104026].
   
  \bibitem{Damour:2004kw} 
  T.~Damour and A.~Vilenkin,
 ``Gravitational radiation from cosmic (super)strings: Bursts, stochastic background, and observational windows",
  Phys.\ Rev.\ D {71}, 063510 (2005)  [hep-th/0410222].
  
 \bibitem{Suresh:2023hkz}
    N.~Suresh and D.~F.~Chernoff,
    ``Modeling the Beam of Gravitational Radiation from a Cosmic String Loop",
    arXiv:2310.00825 [astro-ph.CO].
    
    \bibitem{Leblond:2009fq} 
  L.~Leblond, B.~Shlaer and X.~Siemens,
  ``Gravitational Waves from Broken Cosmic Strings: The Bursts and the Beads",
  Phys. Rev. D {79}, 123519 (2009) [0903.4686].
   
\bibitem{Chernoff:2007pd} 
  D.~F.~Chernoff and S.-H.~H.~Tye,
  ``Cosmic String Detection via Microlensing of Stars",
  arXiv:0709.1139 [astro-ph].

 \bibitem{Hu:2000ke}
 W.~Hu, R.~Barkana, and A.~Gruzinov, 
 ``Cold and fuzzy dark matter", 
 Phys. Rev. Lett. 85, 1158 (2000)  [astro-ph/0003365].
 
\bibitem{Schive:2014dra}
 H.-Y.~Schive, T.~Chiueh, and T.~Broadhurst, 
 ``Cosmic Structure as the Quantum Interference of a Coherent Dark Wave", 
 Nature Phys. 10, 496 (2014) [arXiv:1406.6586].
 
 \bibitem{Hui:2016ltb}
 L.~Hui, J.~P.~Ostriker, S.~Tremaine, and E.~Witten, 
 ``Ultralight scalars as cosmological dark matter", 
 Phys. Rev. D95, 4, 043541 (2017) [arXiv:1610.08297]. 
 
 \bibitem{Luu:2018afg}
 H.~N.~Luu, S.-H.~H.~Tye and T.~Broadhurst,
 ``Multiple Ultralight Axionic Wave Dark Matter and Astronomical Structures",
 Phys. Dark Univ. 30, 100636 (2020) [1811.03771].
 
 \bibitem{Fung:2021wbz}
 L.~W.~H.~Fung, L.~Li, T.~Liu, H.~N.~Luu, and Y.-C.~Qiu, and S-H.~H.~Tye,
 ``Axi-Higgs Cosmology",
 JCAP 08, 057 (2021) [2102.11257].
 
 \bibitem{Sen:1998sm}
   A.~Sen, 
    ``Tachyon condensation on the brane anti-brane system",
    JHEP 08, 012 (1998) [hep-th/9805170].

 \bibitem{Shiu:2002xp}
    G.~Shiu, S.-H.~H.~Tye and I.~Wasserman,
    ``Rolling tachyon in brane world cosmology from superstring field theory",
    Phys. Rev. D67, 083517 (2003) [hep-th/0207119].

\bibitem{Lambert:2003zr}
    N.~D.~Lambert, H.~Liu and J.~M.~Maldacena,
    ``Closed strings from decaying D-branes",
    JHEP 03, 014 (2007) [hep-th/0303139].  
    
 \bibitem{Chen:2003xq}
    X.~Chen,
    ``One loop evolution in rolling tachyon",
    Phys. Rev. D70, 086001 (2004) [hep-th/0311179].

    \bibitem{Sen:2004nf}
    A.~Sen,
   ``Tachyon dynamics in open string theory",
   Int. J. Mod. Phys. A20, 5513 (2005) [hep-th/0410103].

      \bibitem{Leblond:2006cc}
    L.~Leblond and S.~Shandera,
    ``Cosmology of the Tachyon in Brane Inflation",
     JCAP 01, 009 (2007) [hep-th/0610321].

 \bibitem{Brandenberger:2007ca}
    R.~Brandenberger, A.~R.~Frey and L.~C.~Lorenz,
    ``Entropy fluctuations in brane inflation models",
    Int. J. Mod. Phys. A24, 4327 (2009) [0712.2178].

\bibitem{Battefeld:2010rf}
    D.~Battefeld, T.~Battefeld,  H.~Firouzjahi and N.~Khosravi,
    ``Brane Annihilations during Inflation",
      JCAP 07, 009 (2010) [1004.1417].
      
 \bibitem{Banks:1995ch}
    T.~Banks and L.~Susskind,
    ``Brane - anti-brane forces",
    arXiv:hep-th/9511194.
        
           \bibitem{Guth:1982pn}
    A.~H.~Guth and E.~J.~Weinberg,
    ``Could the Universe Have Recovered from a Slow First Order Phase Transition?",
    Nucl. Phys. B212, 321 (1983).
    
    \bibitem{Coleman:1977}
    S.~R.~Coleman,
     ``The Fate Of The False Vacuum. 1. Semiclassical Theory",
     Phys. Rev. D15, 2929 (1977) [Erratum-ibid. D16, 1248 (1977)].
    
   \bibitem{Coleman:1980aw}
  S.~R.~Coleman and F.~De~Luccia,
    ``Gravitational Effects on and of Vacuum Decay",
     Phys. Rev. D21, 3305 (1980).

    \bibitem{Huang:2008jr}
    Q.-G.~Huang and S.-H.~H.~Tye,
    ``The Cosmological Constant Problem and Inflation in the String Landscape",
    Int. J. Mod. Phys. A24, 1925 (2009) [0803.0663].

\bibitem{Silk:1986vc}
   J.~Silk and M.~S.~Turner,
    ``Double Inflation",
   Phys. Rev. D35, 419 (1987).

\bibitem{Hawking:1981fz}
    S.~W.~Hawking and I.~G.~Moss,
    ``Supercooled Phase Transitions in the Very Early Universe",
   Phys. Lett. B110, 35 (1982).
   
\bibitem{Jones:2003ae}
    N.~T.~Jones, L.~Leblond and S.-H.~H.~Tye,
    ``Adding a brane to the brane anti-brane action in BSFT",
   JHEP 10, 002 (2003) [hep-th/0307086].

\bibitem{NANOGrav:2023gor}
    G.~Agazie and others,
    ``The NANOGrav 15 yr Data Set: Evidence for a Gravitational-wave Background",
    Astrophys. J. Lett. 951, 1, L8 (2023) [2306.16213].

\bibitem{NANOGrav:2023hde}
  G.~Agazie and others,
    ``The NANOGrav 15 yr Data Set: Observations and Timing of 68 Millisecond Pulsars",
      Astrophys. J. Lett. 951, 1, L9 (2023) [2306.16217].

\bibitem{Xu:2023wog}
    H.~Xu and others,
    ``Searching for the Nano-Hertz Stochastic Gravitational Wave Background with the Chinese Pulsar Timing Array Data Release I",
    Res. Astron. Astrophys. 23, 7, 075024 (2023) [2306.16216].

\bibitem{EPTA:2023sfo}
    J.~Antoniadis and others,
    ``The second data release from the European Pulsar Timing Array I. The dataset and timing analysis",
    arXiv:2306.16224 [astro-ph.HE].

\bibitem{Zic:2023gta}
    A.~Zic and others,
    ``The Parkes Pulsar Timing Array Third Data Release",
    arXiv:2306.16230 [astro-ph.HE].
   
\bibitem{Ellis:2023oxs}
   J.~Ellis, M.~Fairbairn, G.~Franciolini, G.~H\"utsi, A.~Iovino, M.~Lewicki, M.~Raidal, J.~Urrutia, V.~Vaskonen, H.~and Veerm\"ae,
    ``What is the source of the PTA GW signal?",
        arXiv:2308.08546 [astro-ph.CO].
   
   \bibitem{Witten:1984rs}
    E.~Witten,
    ``Cosmic Separation of Phases",
    Phys. Rev. D30, 272 (1984).
    
   \bibitem{Hogan:1986qda}
    C.~J.~Hogan,
    ``Gravitational radiation from cosmological phase transitions",
    Mon. Not. Roy. Astron. Soc. 218, 629 (1986).
    
    \bibitem{Turner:1990rc}
   M.~S.~Turner and F.~Wilczek,
    ``Relic gravitational waves and extended inflation",
    Phys. Rev. Lett. 65, 3080 (1990).
    
   \bibitem{Kamionkowski:1993fg}
    M.~Kamionkowski, A.~Kosowsky and M.~S.~Turner,
    ``Gravitational radiation from first order phase transitions",
   Phys. Rev. D49, 2837 (1994) [astro-ph/9310044].
    
    \bibitem{Chialva:2010jt}
    D.~Chialva,
    ``Gravitational waves from first order phase transitions during inflation",
    Phys. Rev. D83, 023512 (2011) [1004.2051].
    
    \bibitem{Hindmarsh:2015qta}
    M.~Hindmarsh, S.~J.~Huber, K.~Rummukainen and D.~J.~Weir,
   ``Numerical simulations of acoustically generated gravitational waves at a first order phase transition",
    Phys. Rev. D92, 12, 123009 (2015) [1504.03291].
    
   \bibitem{Jiang:2015qor}
   H.~Jiang, T.~Liu, S.~Sun and Y.~Wang,
    ``Echoes of Inflationary First-Order Phase Transitions in the CMB",
    Phys. Lett. B765, 339 (2017) [1512.07538]

\bibitem{An:2020fff}
    H.~An, K.-F.~Lyu, L.-T.~Wang and S.~Zhou,
    ``A unique gravitational wave signal from phase transition during inflation",
    Chin. Phys. C46, 10, 101001 (2022) [2009.12381]

   \bibitem{Alishahiha:2004eh}
    M.~Alishahiha, E.~Silverstein and D.~Tong,
    ``DBI in the sky",
    Phys. Rev. D70, 123505 (2004) [hep-th/0404084]

   \bibitem{Shandera:2006ax}
    S.~E.~Shandera and S.-H.~H.~Tye,
    ``Observing brane inflation",
       JCAP 05, 007 (2006) [hep-th/0601099]

\end{thebibliography}
\end{document}